\magnification=1200
\overfullrule=0pt
\baselineskip12pt





\def\hexnumber@#1{\ifcase#1 0\or 1\or 2\or 3\or 4\or 5\or 6\or 7\or 8\or
 9\or A\or B\or C\or D\or E\or F\fi}

\def\bno{\bigskip\noindent}

\def\today{\noindent\number\day 
\space\ifcase\month
 \or January
 \or February
 \or March
 \or April
 \or May
 \or June
 \or July
 \or August
 \or September
 \or October
 \or November
 \or December
\fi \space\number\year}


\def\Bbb{\bf}  
\def\R  {{\Bbb R}} 
\def\Z  {{\Bbb Z}}

\def\E  {{\cal E}}

 
\def\b {\beta} 
\def\g {\gamma}
\def\tg {\tilde \gamma}

\def\s {\sigma}
\def\bs {\bar \sigma}
\def\t {\tau}

\def\L {\Lambda} 
\def\l {\lambda}

\def\O {\Omega} 
 
\def\D{\Delta}

\def\p{\partial}

\def\V{\raise-.6ex\hbox{$\scriptstyle V$}}
\def\L{\raise-.6ex\hbox{$\scriptstyle L$}}
\def\Vc{\raise-.6ex\hbox{$\scriptstyle V^c$}}



\centerline{\bf ON THE UNIQUENESS OF GIBBS STATES IN THE}
\centerline{\bf PIROGOV-SINAI THEORY$^*$}
\bigskip
\centerline{\bf J.L. Lebowitz$^{a},\;$ and A.E. Mazel$^{a,b}$}
\footnote{} {$^a$\quad Department of Mathematics and Physics, Rutgers
University, New Brunswick, NJ 08903, USA} 
\footnote{}{$^b$\quad International Institute of Earthquake Prediction
Theory and Mathematical Geophysics, Russian Academy of Sciences, Moscow
113556, RUSSIA}
\footnote{} {$^*$\quad Work supported by NSF grant DMR 92-13424}

\bigskip
\bigskip\noindent
{\bf Abstract.} We prove that, for low-temperature systems considered in
the Pirogov-Sinai theory, uniqueness in the class of translation-periodic
Gibbs states implies global uniqueness, i.e. the absence of any
non-periodic Gibbs state. The approach to this infinite volume state is
exponentially fast. 

\bigskip\noindent
{\bf Key words:} Gibbs state, Pirogov-Sinai theory, uniqueness, cluster
(polymer) expansion.
\bigskip
\centerline{\bf Dedicated to the memory of Roland Dobrushin}
\bno
{\bf 1. Introduction}

\medskip
The problem of uniqueness of Gibbs states was one of R.L.Dobrushin's
favorite subjects in which he obtained many classical results. In
particular when two or more transla\-ti\-on-periodic states coexist, it is
natural to ask whether there might also exist other, non
translation-periodic, Gibbs states, which approach asymptotically, in
different spatial directions, the translation periodic ones. The
affirmative answer to this question was given by R.L.Dobrushin with his
famous construction of such states for the Ising model, using $\pm$
boundary conditions, in three and higher dimensions [D]. Here we consider
the opposite situation: we will prove that in the regions of the
low-temperature phase diagram where there is a unique translation-periodic
Gibbs state one actually has {\it global uniqueness} of the limit Gibbs
state. Moreover we show that, uniformly in boundary conditions, the finite
volume probability of any local event tends to its infinite volume limit
value exponentially fast in the diameter of the domain.

The first results concerning this problem in the framework of the
Pirogov-Sinai theory [PS] were obtained by R.L.Dobrushin and E.A.Pecherski
in [DP]. The Pirogov-Sinai theory describes the low-temperature phase
diagram of a wide class of spin lattice models, i.e. it determines all
their translation-periodic limit Gibbs states [PS], [Z].  The results of
[DP], corrected and extended in [Sh], imply that, for any values of
parameters at which the model has a unique ground state, the Gibbs state
is unique for sufficiently small temperatures. But the closer these
parameters are to the points with non unique ground state the smaller the
temperature for which uniqueness of the Gibbs state is given by this
method.  Independently, an alternative method leading to similar results
was developed in [M1,2].

The main difficulty in establishing the results of this type is due to the
necessity of having sufficiently detailed knowledge of the partition
function in a finite domain with an {\it arbitrary} boundary
condition. This usually requires a detailed analysis of the geometry of
the so called {\it boundary layer} produced by such a boundary condition
(see [M1,2], [Sh]). Here we develop a new simplified approach to the
problem. The simplification is achieved by transforming questions
concerning the finite volume Gibbs measure with an arbitrary boundary
conditions into questions concerning the distribution with a {\it stable}
(in the sense of [Z]) boundary condition. The latter can be easily
investigated by means of the {\it polymer expansion} constructed for it in
the Pirogov-Sinai theory. This also allows the extension of the uniqueness
results from systems with a unique ground state to the case with several
ground states but unique {\it stable ground state} (see [Z]).

Since the publication of the paper [PS] about twenty years ago the
Pirogov-Sinai theory was extended in different directions. For a good
exposition of the initial theory we refer the reader to [Si] and [Sl]. Some
of the generalizations can be found in [BKL], [BS], [DS], [DZ], [HKZ] and
[P]. Below we present our results in the standard settings of [PS] and [Z]:
a {\it finite spin space with a translation-periodic finite potential of
finite range}, a {\it finite degeneracy of the ground state} and a {\it
stability of the ground states} expressed via the so called {\it Peierls} or
{\it Gertzik-Pirogov-Sinai condition}. The extension to other cases is
straightforward. Our method also works for unbounded spins, see [LM].

\bigskip
\bno
{\bf 2. Models and Results.}

\medskip
The models are defined on some lattice, which for the sake of simplicity
we take to be the $d$-dimensional ($d \ge 2$) cubic lattice $\Z^d$. The
spin variable $\s_x$ associated with the lattice site $x$ takes 
values from the finite set $S=\{1,2,\ldots, |S|\}$. The energy of the
configuration $\s \in S^{\Z^d}$ is given by the formal Hamiltonian
$$H_0(\s)=\sum_{A \subset \Z^d,\; {\rm diam}\;A \le r }
U_0(\s_A). \eqno{(1)}$$ 
Here $\s_A \in S^A$ is a configuration in $A \subset \Z^d$ and the
potential $U_0(\s_A): \;S^A \mapsto \R$, satisfies
$U_0(\s_A)=U_0(\s_{A+y})$ for any $y$ belonging to some subgroup of $\Z^d$
of finite index and the sum is extended over subsets $A$ of $\Z^d$ with a
diameter not exceeding $r$. Accordingly for a finite domain $V \subset
\Z^d$, with the boundary condition $\bs_{\Vc}$ given on its complement
$V^c= \Z^d \setminus V$, the conditional Hamiltonian is
$$H_0(\s_{\V}|\bs_{\Vc})=\sum_{A \cap V \not= \emptyset,\; {\rm diam}\;A
\le r } U_0(\s_A), \eqno{(2)}$$
where $\s_A=\s_{A \cap V} + \bs_{A \cap V^c}$ for $A \cap V^c \not=
\emptyset$, i.e. the spin at site $x$ is equal to $\s_x$ for $x \in A \cap
V$ and $\bs_x$ for $x \in A \cap V^c$.

A {\it ground state} of (1) is a configuration $\s$ in $\Z^d$ whose energy
cannot be lowered by changing $\s$ in some local region. We assume that (1)
has a finite number of translation-periodic (i.e. invariant under the
action of some subgroup of $\Z^d$ of finite index) ground states. By a
standard trick of partitioning the lattice into disjoint cubes $Q(y)$
centered at $y \in q \Z^d$ with an appropriate $q$ and enlarging the spin
space from $S$ to $S^{Q}$ one can transform the model above into a model
on $q\Z^d$ with a {\it translation-invariant} potential and only {\it
translation-invariant or non-periodic} ground states. Hence, without loss
of generality, we assume translation-invariance instead of
translation-periodicity and we permute the spin so that the ground states
of the model will be $\s^{(1)}, \ldots, \s^{(m)}$ with $\s^{(k)}_x=k$ for
any $x \in \Z^d$. Taking $q >r$ one obtains a model with nearest neighbor
and next nearest neighbor (diagonal) interaction, i.e. the potential is
not vanishing only on lattice cubes $Q_1$ of linear size 1, containing
$2^d$ sites.

Given a configuration $\s$ in ${\Z^d}$ we say that site $x$ {\it is in the
$k$-th phase} if this configuration coincides with $\s^{(k)}$ inside the
lattice cube $Q_2(x)$ of linear size 2 centered at $x$. Every connected
component of sites not in one of the phases is called a {\it contour of
the configuration $\s$}. It is clear that for $\s =\s_{\V} +
\s^{(k)}_{\Vc}$ contours are connected subsets of $V$ which we denote by
$\tg_1(\s), \ldots, \tg_l(\s)$. The important observation is that the
excess energy of a configuration $\s$ with respect to the energy of the
ground state $\s^{(k)}$ is concentrated along the contours of $\s$. More
precisely,
$$H_0(\s_{\V}|\s^{(k)}_{\Vc}) - H_0(\s^{(k)}_{\V}|\s^{(k)}_{\Vc})=
\sum_{i} H_0(\tg_i(\s)), \eqno{(3)}$$
where
$$H_0(\tg_i(\s))=\sum_{Q_1:\; Q_1 \subseteq \tg_i(\s)}
\left( U_0(\s_{Q_1})-U_0(\s^{(k)}_{Q_1}) \right) \eqno{(4)}$$
and the sum is taken over the unit lattice cubes $Q_1$, containing $2^d$
sites. The Peierls condition is
$$H_0(\tg_i(\s)) \ge \t |\tg_i(\s)|, \eqno{(5)}$$
where $\t >0$ is an absolute constant and $|\tg_i(\s)|$ denotes the
number of sites in $\tg_i(\s)$.

Consider now a family of Hamiltonians $H_n(\s)=\displaystyle{\sum_{A
\subset \Z^d,\;{\rm diam}\; A \le r}} U_n(\s_A)$, $n=1, \ldots,m-1$
satisfying the same conditions as $H_0$ with the same or smaller set of
translation-periodic ground states. For $\l=(\l_1, \ldots, \l_{m-1})$
belonging to a neighborhood of the origin in $\R^{m-1}$ define a perturbed
formal Hamiltonian
$$H=H_0+\sum_{n=1}^{m-1} \l_n H_n. \eqno{(6)}$$
Here $\l_n H_n$ play the role of generalized magnetic fields removing the
degeneracy of the ground state. The finite volume Gibbs distribution is
$$\mu_{\V,\bs_{\Vc}}(\s_{\V})={\exp \left[-\b H(\s_{\V}|\bs_{\Vc}) \right]
\over \Xi(V|\bs_{\Vc})}, \eqno{(7)}$$
where $\b >0$ is the inverse temperature and $\mu_{\V,\bs_{\Vc}}(\s_{\V})$ is
the probability of the event that the configuration in $V$ is $\s_{\V}$,
given $\bs_{\Vc}$. Here the conditional Hamiltonian is
$$H(\s_{\V}|\bs_{\Vc})= \sum_{Q_1:\;Q_1 \cap V \not= \emptyset}
U(\s_{Q_1}), \quad U(\cdot)=\sum_{n=0}^{m-1}
U_n(\cdot)  \eqno{(8)}$$
and the partition function is
$$\Xi(V|\bs_{\Vc})=\sum_{\s_{\V}} \exp \left[-\b H(\s_{\V}|\bs_{\Vc}) \right]
\eqno{(9)}$$ 

The notion of a {\it stable ground state} was introduced in [Z] (see also
the next section) and it is crucial for the Pirogov-Sinai theory because
of the following theorem.

\medskip\noindent
{\bf Theorem [PS], [Z].} {\sl Consider a Hamiltonian $H$ of the form (6)
satisfying all the conditions above. Then for $\b$ large enough, $\b \ge
\b_0(\l)$, every stable ground state $\s^{(k)}$ generates a
translation-invariant Gibbs state
$$\mu^{(k)}(\cdot) = \lim_{V \to \Z^d}
\mu_{\V,\s^{(k)}_{\Vc}}(\cdot). \eqno{(10)}$$
These Gibbs states are different for different $k$ and they are the only
translation-periodic Gibbs states of the system.}

{\bf Remark.} Given $H_n,\; n=0,\ldots,m-1$ there exists sufficiently
small $\l_0$ such that for $\displaystyle \max_{1 \le n \le m-1} |\l_n|
\le \l_0$ the quantity $\b_0$ becomes independent on $\l$.

\medskip
An obvious corollary of the above theorem is

\medskip\noindent
{\bf Corollary.} {\sl If there is only a single stable ground state, say
$\s^{(1)}$, then for $\b \ge \b_0(\l)$ there is a unique
translation-periodic Gibbs state 
$$\mu^{(1)}(\cdot) = \lim_{V \to \Z^d}
\mu_{\V,\s^{(1)}_{\Vc}}(\cdot). \eqno{(11)}$$} 

\medskip
Our extension of this result is given by

\medskip\noindent
{\bf Theorem.} {\sl Consider a Hamiltonian $H$ of the form (6) satisfying
all the conditions above. Suppose that given $\l$ and $\b \ge \b_0(\l)$
there exist a single stable ground state, say $\s^{(1)}$. Then the Gibbs
state $\mu(\cdot)=\mu^{(1)}(\cdot)$ is unique. Moreover, for finite $A
\subset V$, such that ${\rm dist}(A, V^c) \ge 2d\; {\rm diam}A +
C_1(\t,\b,\l,d)$, any con\-fi\-gu\-ration $\s_A$ and any boundary
condition $\bs_{\Vc}$, one has 
$$\left| \mu_{\V,\bs_{\Vc}}(\s_A) - \mu(\s_A) \right| \le \exp \left[ -
C_2(\t, \b, \l, d) \;{\rm dist}\; (A,V^c) \right], \eqno{(12)}$$ where
$C_1, C_2 > 0$.}

\medskip
{\bf Remark.} In contrast with [DP], [Sh] and [M1,2] the theorem above
treats the situation when there are several ground states with only one of
them being stable. Moreover, the result is true for all sufficiently low
temperatures not depending on how close the parameters are to the points
with non unique ground state. If some of the conditions of the Theorem are
violated the statement can be wrong.

The simplest counterexample can be constructed from the Ising model in
$d=3$. It is well-known that at low temperatures this model contains
precisely two translation-invariant Gibbs states taken into each other by
$\pm$ symmetry and infinitely many non translation-invariant Gibbs states,
i.e. the Dobrushin states mentioned earlier. Identifying configurations
taken into each other by $\pm$ symmetry one obtains a model with unique
translation-invariant Gibbs state and infinitely many
non translation-invariant ones. The condition of the Theorem which is not
true for this factorized model is the finiteness of the potential: the
model contains a hard-core constraint.

Formally speaking, one can consider a model with spin variables still
taking values $\pm 1$ but assigned to bonds of the lattice. Then the
Hamiltonian is the sum of the spins over all lattice bonds multiplied by a
negative coupling constant. The hard-core constraint says that the product
of spins along any lattice plaquette is 1.

Another example based on a gauge model can be found in [B]. For this model
the spin space is finite, the potential is finite and of finite radius but
the Peierls condition is violated and the number of ground states is
infinite. On the other hand, after a proper factorization, this system can
be transformed into a model with a hard-core restriction similar to that
discussed above.

It is known [LML] that for systems satisfying FKG inequalities the
uniqueness of the translation-invariant Gibbs state implies uniqueness of all Gibbs states at any temperature. This makes it tempting to assume that the conclusions of our theorem hold beyond the low temperature region covered by the Pirogov-Sinai theory.  We are not aware of any counterexample.

\bigskip
\bno
{\bf 3. Proof of the Theorem}

\medskip
In this section we assume that the reader is familiar with the
Pirogov-Sinai theory and we only list the appropriate notation. Then
we quote some necessary results and proceed to the proof of the theorem.

\bigskip
\bno
{\bf Preliminaries.} 
A {\it contour} is a pair $\g=(\tg, \s_{\tg})$ consisting of the {\it
support} $\tg$ and the configuration $\s_{\tg}$ in it. The components of the
{\it interior} of the contour $\g$ are denoted by $Int_j \g$ and the {\it
exterior} of $\g$ is denoted by $Ext \g$. The family $\{\tg, Int_j \g, Ext
\g\}$ is a partition of $\Z^d$. The configuration $\s_{\tg}$ can be uniquely
extended to the configuration $\s'$ in $\Z^d$ taking constant values
$I_j(\g)$ and $E(\g)$ on the connected components of $\tg^c$. Generally
these values are different for different components. The contour $\g$ is
said to be {\it from the phase $k$} if $E(\g)=k$. The {\it energy} of the
contour $\g$ is 
$$H(\tg(\s))=\sum_{Q_1:\; Q_1 \subseteq \tg(\s)} \left(
U(\s_{Q_1})-U(\s^{(E(\g))}_{Q_1}) \right). \eqno{(13)}$$

The {\it statistical weight} of $\g$ is
$$w(\g)=\exp (-\b H(\g)) \eqno{(14)}$$
and satisfies
$$0 \le w(\g) \le e^{- \b \t |\tg|}. \eqno{(15)}$$

The {\it renormalized statistical weight} of the contour is
$$W(\g)=w(\g)\prod_j { \Xi(Int^*_j \g |I_j(\g)) \over 
\Xi(Int^*_j \g |E(\g))}, \eqno{(16)}$$
where for any $A \subset \Z^d$ we denote
$$A^*=\{ x\in A |\; x \;\;{\rm is\ not\ adjacent\ to\ } A^c\}. \eqno{(17)}$$

The contour $\g$ is {\it stable} if
$$W(\g) \le \exp \left[-{1 \over 3}\b \t |\tg| \right] \eqno{(18)}$$
and the ground state $\s^{(k)}$ is {\it stable} if all contours $\g$ with
$E(\g)=k$ are stable. It is known (see [Z]) that at least one of the
ground states is stable. Because of (18) for any $x \in \Z^d$, $N\ge 1$
and $\b$ large enough
$$\sum_{\g:\; (Ext \g)^c \ni x,\; |\tg| \ge N} W(\g) \le e^{-C_3 N},
\eqno{(19)}$$
where $C_3=C_3(\t, \b, d)$ is positive and monotone increasing in $\t$ and
$\b$. In particular 
$$\sum_{\g:\; (Ext \g)^c \ni x} W(\g) \le C_4, \eqno{(20)}$$
where $C_4=e^{-C_3}$. For the stable ground state $\s^{(k)}$ the
corresponding partition function can be represented as
$$\Xi(V|k)=e^{-\b H(\s^{(k)}_{\V} | \s^{(k)}_{\Vc})}
\sum_{[\g_i]^s\in V,\; E([\g_i]^s)=k} \prod_i W(\g_i), \eqno{(21)}$$
where the sum is taken over all collections of contours $[\g_i]$ such that
$\tg_i$ are disjoint, $E(\g_i)=k$ for all $i$ and $\tg_i \subseteq V$ for
all $i$.

Representation (21) and estimate (18) allow to write an absolutely
convergent polymer expansion
$$\log \Xi(V|k)=-\b H(\s^{(k)}_{\V} | \s^{(k)}_{\Vc}) +\sum_{\pi^{(k)} \in V}
W(\pi^{(k)}), \eqno{(22)}$$
where the sum is taken over so called {\it polymers} $\pi^{(k)}$ of the
phase $k$ belonging to the domain $V$. By definition a polymer
$\pi^{(k)}=(\g_i)$ is a collection of, not necessarily different, contours
$\g_i$ of the phase $k$ such that $\cup_i \tg_i$ is connected. The
statistical weight $W(\pi^{(k)})$ is uniquely defined via $W(\g_i)$ and
satisfies the estimate (see [Se])
$$|W(\pi^{(k)})| \le \exp \left[ -\left({1 \over 3} \b \t - 6d
\right)\sum_i|\tg_i| \right] \eqno{(23)}$$
implying
$$\sum_{\pi^{(k)}=(\g_i):\; \cup_i(Ext \g_i)^c \ni x,\; \sum_i |\tg_i| \ge N}
|W(\pi^{(k)})| \le e^{-C_3 N}. \eqno{(24)}$$

Denote by $\mu^{(k)}_{\V}(\{\g_i\}, ext)$ the probability of the event
that all contours of the collection $\{\g_i\}$ are external ones inside
$V$. By the construction
$$\mu^{(k)}_{\V}(\{\g_i\}, ext) \le \prod_i W(\g_i).  \eqno{(25)}$$ 
     From the polymer expansion (22) and estimate (24) it is not hard to
conclude that for $\{\g_i\}$ with dist$(\cup_i \tg_i, V^c) \ge |\cup_i
\tg_i|$
$$\left|\mu^{(k)}_{\V}(\{\g_i\}, ext) - \mu^{(k)}(\{\g_i\}, ext) \right| \le
\mu^{(k)}(\{\g_i\}, ext) |\cup_i \tg_i| \exp \left[ -C_5\; {\rm dist\;}
(\cup_i \tg_i, V^c) \right], \eqno{(26)}$$
where $C_5=C_5(\t, \b, d)$ is positive and monotone increasing in $\t$ and
$\b$. For any $A \in V$, dist$(A,V^c) \ge$ diam$A$, and any $\s_A$
estimate (26) implies in a standard way that
$$\left|\mu^{(k)}_{\V}(\s_A) - \mu^{(k)}(\s_A) \right| \le \exp \left[ -C_6\;
{\rm dist\;} (A, V^c) \right], \eqno{(27)}$$
where again $C_6=C_6(\t,\b,d)$ is positive and monotone increasing in $\t$
and $\b$. 

For an arbitrary boundary condition $\bs_{\Vc}$ the probability
$\mu_{\V,\bs_{\Vc}}(\s_A)$ depends only on $\bs_{\atop \p V}$, where
$$\p V=\{x \in V^c:\; x \;{\rm is\ adjacent \ to\ } V\}, \eqno{(28)}$$ 
and we freely use the notation $\mu_{\V,\bs_{\atop \p V}}(\s_A)$.
   From now on we suppose that $\s^{(1)}$ is the only stable ground state
of $H$ and denote by $\mu_{\V}(\cdot)$ the Gibbs distributions with the
stable boundary condition $\s^{(1)}_{\Vc}$. 

For a domain $V$ with the boundary condition $\s^{(1)}_{\Vc}$ fix $l \le
L$ sites belonging to $\p V$ and consider a collection $\{\g_i\}^e \in V$
of external contours  touching $\p V$ at one of these sites. A smaller
domain $V'=\cup_i \cup_j Int_j \g_i$ has a natural boundary condition
$\s'_{\p V'}$ induced by $\cup_i \s_{\tg_i}$. Given $M \ge \sum_i
|\tg_i|$ denote by $\E_{\{\g_i\}^e, M}$ the event that the total number of
sites in all adjacent to $\p V'$ connected components of sites not in the
1-st phase is not less than $M - \sum_i |\tg_i|$. According to the Theorem
of Section~3.2 in [Z]
$$\mu_{\V',\s'_{\p V'}}\left( \E_{\{\g_i\}^e, M} \right) \le e^{
-C_7(\t,\b,\l,d)( M - \sum_i |\tg_i|) + C_8(\t,\b,\l,d)\sum_i |\tg_i|}.
\eqno{(29)}$$
The positive constants $C_7$ and $C_8$ tend to $0$ as $\b \to \infty$ or
$(\b, \l)$ approaches the manifold on which $\s^{(1)}$ is not the only
stable ground state. For different $\{\g_i\}^e$ the events
$\E_{\{\g_i\}^e, M}$ are disjoint and for their union
$\E_{V,M,L}=\cup_{\{\g_i\}^e \in V} \E_{\{\g_i\}^e, M}$ one has the estimate
$$\eqalignno{
\mu_{\V} \left(\E_{V,M,L}\right) &= \sum_{\{\g_i\}^e \in V}
\mu_{\V',\s'_{\p V'}}\left( \E_{\{\g_i\}^e, M} \right) \mu_{\V}
\left(\{\g_i\}^e \right) \cr
&\le \sum_{\{\g_i\}^e \in V} e^{-C_7 (M - \sum_i |\tg_i|) +C_8 |\tg_i|}
\prod_i W(\g_i) \cr
&\le e^{-C_7 M} \sum_{\{\g_i\}^e \in V} \prod_i 
e^{ -\left({1 \over 3} \t \b - C_7 -C_8 \right) |\tg_i|} \cr
&\le  e^{-C_7 M} (1+C_4)^L \cr
&\le e^{-C_7 M + C_4 L}. &(30)\cr}$$
Finally observe that for any $A \subseteq V$ and any $\s_A$
$$\mu_{\V}(\s_A)e^{- C_9 L(\bs_{\p V})} \le \mu_{\V,\bs_{\Vc}}(\s_A) \le
\mu_{\V}(\s_A)e^{ C_9 L(\bs_{\p V})}, \eqno{(31)}$$
where $L(\bs_{\p V})$ is the number of sites $x \in \p V$ with $\bs_{\p V}
\not= 1$ and 
$$C_9=2^d \b \max_{\s_{Q_1}} \left| U(\s_{Q_1})\right|. \eqno{(32)}$$

\bno {\bf Proof.}  We are now ready to prove the theorem. Take an integer
$N>0$ and suppose that $V$ contains a cube $Q_{6N}$ with sides of length
6N centered at the origin. From now on all cubes are assumed to be
centered at the origin. Let $Q_{N'},\; N'\ge 6N$ be the maximal cube
contained in $V$. Denote $\p'V=\p Q_{N'} \cap \p V$. First we consider
boundary conditions $\bs_{\atop \p V}$ which coincide with $\s^{(1)}$ on
$\p V \setminus \p'V$ and differ from $\s^{(1)}$ on $\p' V$ by at most
$\sqrt{N}$ lattice sites. 

Given $\s_{\V}$ denote by $\O(\s_{\V})$ the union of the connected
components of the set $\{x \in V:\; x$~is not in the 1-st phase~$\}$
adjacent to $\{x \in \p V:\; \bs_x \not= 1\}$. This set is called the {\it
boundary layer} of $\s_{\V}$ and we denote by $\O_i(\s_{\V})$ its
connected components.  
Introduce the event $\E_0=\{\s_{\V}:\; \O(\s_{\V}) \cap Q_{4N} \not=
\emptyset \}$. By construction for $\s_{\V} \in \E_0$ every
$\O_i(\s_{\V})$ touches $\p V$ at some site $x \in \p V$ with $\bs_x
\not=1$ and there exists at least one component $\O_i$ intersecting
$Q_{4N}$. Without loss of generality we suppose that it is
$\O_1(\s_{\V})$. This leads to the estimate
\goodbreak
$$\eqalignno{
\mu_{\V,\bs_{\Vc}}(\E_0) &\le
e^{ C_9 \sqrt{N}} \mu_{\V}(\E_0) \cr
&\le e^{ C_9 \sqrt{N}} \sqrt{N} e^{-C_3 N}
\left( 1+ C_4 \right)^{\sqrt{N}} \cr
&\le e^{-C_{10} N} \;. &(33)\cr}$$

In the first inequality of (33) we used (31) reducing the problem to the
calculation for the stable boundary condition $\s^{(1)}$. The second
inequality comes in a standard way from the cluster expansion for
$\mu_{\V}(\cdot)$. Indeed, in the domain $V$ with the stable boundary
condition $\s^{(1)}_{\Vc}$ every component $\O_i$ contains an external
contour $\g_i$ such that $E(\g_i)=1$, $\tg_i \subseteq \O_i$ and $\O_i
\subseteq (Ext(\g_i))^c$. One may simply say that $\g_i$ is the external
boundary of $\O_i$ and clearly $\tg_i$ touches $\p V$. If $\O_1$
intersects $Q_{4N}$ then $\tg_1$ intersects or encloses $Q_{4N}$. The
number of possibilities to chose the site $x \in \p V,\; \bs_x \not=1$ at
which $\tg_1$ touches $\p V$ does not exceed $\sqrt{N}$ which produces the
factor $\sqrt{N}$ in the estimate. The next factor estimates the sum of
the statistical weights of all possible $\g_1$ touching this site. It is
based on (19) and takes into account the fact that the diameter of
$\tg_1$, and hence $|\tg_1|$, is not less than $N$. The constant $C_4$
(see (20)) estimates the sum of the statistical weights of all possible
$\g_i$ touching given lattice site and $\left( 1+ C_4 \right)^{\sqrt{N}}$
estimates the statistical weight of all possibilities to choose $\{\g_i$,
$i \not= 1\}$. The whole estimate uses (25) and the fact that
$\mu_{\V}(\{\g_i\},ext)$ is the upper bound for the sum of
$\mu_{\V}$-probabilities of boundary layers $\O=\{\O_i\}$ having $\g_i$ as
the boundary of $\O_i$. The third inequality in (33) is trivial for
$C_{10}=0.5C_3$ and $N \ge 4 ( C_9 +1 + C_4)^2 C_3^{-2}$.

Denote by $\E_0^c$ the complement of $\E_0$. If $\s_{\V} \in \E_0^c$ then $(V
\setminus \O(\s_{\V}))^* \supseteq Q_{4N-2}$ (see (17) for the definition
of $(\cdot)^*$). It is not hard to see that the
configuration $\s_{\V} \in \E_0^c$ equals 1 on the boundary of $(V \setminus
\O(\s_{\V}))^*$. Now fix $A \subset Q_{2N}$ and $\s_A$. In view of (27) one
has
$$\left| \mu_{\V,\bs_{\Vc}}(\s_A | \E_0^c) - \mu (\s_A) \right| \le e^{-C_6
N}. \eqno{(34)}$$
This gives us
$$\eqalignno{
\left| \mu_{\V,\bs_{\Vc}}(\s_A) - \mu(\s_A) \right| &\le
\left| \mu_{\V,\bs_{\Vc}}(\s_A |\E_0) - \mu(\s_A) \right|
\mu_{\V,\bs_{\Vc}}(\E_0) \cr 
&+\left| \mu_{\V,\bs_{\Vc}}(\s_A |\E_0^c) - \mu(\s_A) \right|
\mu_{\V,\bs_{\Vc}}(\E_0^c) \cr 
&\le e^{-C_{10} N}+ e^{-C_6 N} \cr
&\le e^{-C_{11} N} \;, &(35)\cr}$$
where $C_{11}=0.5 \min (C_{10}, C_6)$ and $N \ge \log 2 /C_{11}$.

To extend (35) to the wider class of boundary conditions we suppose that
$V \supseteq Q_{8N}$ and $Q_{N'},\; N' \ge 8N$ is the maximal cube
contained in $V$. Now we consider boundary conditions $\bs_{\atop \p V}$
which coincide with $\s^{(1)}$ on $\p V \setminus \p'V$ and 
differ from $\s^{(1)}$ on $\p' V$ by at most $(\sqrt{N})^2$
lattice sites. Denote $\D_i=Q_{8N-2i+2} \setminus Q_{8N-2i}$ and let
$\O^{(i)}(\s_{\V})$ be a union of the connected components of the set $\{x
\in V \setminus Q_{8N-2i}:\; \s_x \not= 1\}$ adjacent to $\{x \in \p V:\;
\bs_x \not= 1\},\; i=1, \ldots, N$. Introduce disjoint events
\goodbreak
$$\eqalign{
\E_1=\{\s_{\V}:\; &| \O^{(1)}(\s_{\V}) \cap \D_{1}| < \sqrt{N} \}, \cr
\E_i=\{\s_{\V}:\; &| \O^{(1)}(\s_{\V}) \cap \D_1| \ge \sqrt{N}, \ldots,
| \O^{(i-1)}(\s_{\V}) \cap \D_{i-1}| \ge \sqrt{N}, \cr
&| \O^{(i)}(\s_{\V}) \cap \D_{i}| < \sqrt{N} \}\cr}$$
\centerline{and}
$$\E_c=\left( \bigcup_{i=1}^{N} \E_i \right)^c \;. \eqno{(36)}$$
If $\s_{\V} \in \E_c$ than the boundary contour $\O(\s_{\V})$ contains at
least $N \sqrt{N}$ sites. Hence (30) implies the following estimate
for the probability of $\E_c$
$$\eqalignno{
\mu_{\V,\bs_{\Vc}}(\E_c) &\le 
e^{ C_9 (\sqrt{N})^2} \mu_{\V}(\E_c) \cr
&\le e^{ C_9 (\sqrt{N})^2}  e^{-C_7 N \sqrt{N}+C_4 (\sqrt{N})^2} \cr
&\le e^{-C_{12} N\sqrt{N}} \;, &(37)\cr}$$
where $C_{12}=0.5 C_7$ and $N \ge 4( C_9 +C_4)^2 C_7^{-2}$.

For $\s_{\V} \in \E_i$ consider the volume $V_i =(V \setminus
\O^{(i)}(\s_{\V}))^*\cup Q_{8N-2i}$ with the boundary condition
$\bs_{\Vc}+\s_{V\setminus V_i}$. By construction the number of sites $x
\in \p V_i$ with $\s_x \not= 1$ is less than $\sqrt{N}$ and one can apply
(35) to obtain the bound
$$\left | \mu_{\V,\bs_{\Vc}}(\s_A | \E_i) -\mu(\s_A) \right| \le 
e^{-C_{11} N} \;. \eqno{(38)}$$
Joining (37) and (38) we conclude
$$\eqalignno{
\left | \mu_{\V,\bs_{\Vc}}(\s_A) - \mu(\s_A) \right| &= \left |
\sum_{i=1}^{N} \mu_{\V,\bs_{\Vc}}(\E_i) \bigg ( \mu_{\V,\bs_{\Vc}}(\s_A |
\E_i) -\mu(\s_A) \bigg ) \right. \cr
&+ \left. \phantom{\sum_{i=1}^{N}} \hskip-1.5em
\mu_{\V,\bs_{\Vc}}(\E_c)\bigg ( \mu_{\V,\bs_{\Vc}}(\s_A | \E_c)
-\mu(\s_A) \bigg ) \right| \cr
&\le e^{-C_{11} N}+e^{-C_{12} N\sqrt{N}} \cr
&\le 2e^{-C_{11} N} \;, &(39)\cr}$$
where $N \ge (C_{11}/C_{12})^2$. Expression (39) is a version of (35) which is
weaker by the factor 2 in the RHS but is applicable to the wider class of
boundary conditions containing $(\sqrt{N})^2$ unstable sites instead of
$\sqrt{N}$ for (35).

The argument leading from (35) to (39) can be iterated several times. The
first iteration treats the following situation.  Suppose that $V \supseteq
Q_{10N}$ and $Q_{N'},\; N' \ge 10N$ is the maximal cube contained in
$V$. Consider boundary conditions $\bs_{\atop \p V}$ which coincide with
$\s^{(1)}$ on $\p V \setminus \p'V$ and differ from $\s^{(1)}$ on $\p' V$
by at most $(\sqrt{N})^3$ lattice sites. Then the analogue of (39) is
$$\left| \mu_{\V,\bs_{\Vc}}(\s_A) - \mu(\s_A) \right| \le
3e^{-C_{11} N}. 
\eqno{(40)}$$
Similarly after $2d$ iterations one obtains that for any $V \supseteq
Q_{2dN}$ with the boundary condition $\bs_{\atop \p V}$ containing not
more than $(\sqrt{N})^{2d}$ unstable sites
$$\left| \mu_{\V,\bs_{\Vc}}(\s_A) - \mu(\s_A) \right| \le
2de^{-C_{11} N}. 
\eqno{(41)}$$
Now set $C_1=2d \max \left(4 (C_9 +1 + C_4)^2 C_3^{-2},\; \log 2
/C_{11},\; 4(C_9 +C_4)^2 C_7^{-2},\; (C_{11}/C_{12})^2 \right)$ and for
any $A$ and $\s_A$ consider a cube $Q_{2L}$ with $L \ge C_1$ and dist$(A,
Q^c_{2L}) \ge (1 -1/d)L$. Taking $N=L/d$ and $C_2=C_{11} /2d$ for
$V=Q_{2L}$ one obtains (12) from (41). For any $V \supseteq Q_{2L}$ with
$\p V \cap \p Q_{2L} \not= \emptyset$ we have
$$\eqalignno{
\left| \mu_{\V,\bs_{\Vc}}(\s_A) - \mu(\s_A) \right| &\le \sum_{\s_{\p
Q_{2L}}} \left| \mu_{Q_{2L},\s_{\p Q_{2L}}}(\s_A) - \mu(\s_A) \right|
\mu_{\V,\bs_{\Vc}}(\s_{\p Q_{2L}}) \cr
&\le \exp \left[ - C_2 \;{\rm dist}\; (A,V^c) \right], &(42)\cr}$$
which finishes the proof of the {\bf Theorem}.

\medskip
{\bf Acknowledgments:} We are greatly indebted to F.Cesi,
F.Martinelli and S.Shlos\-man for many very helpful discussions.

\bigskip
\goodbreak
\centerline{\bf REFERENCES }
\bigskip

\item{[B]} C.Borgs, ``Translation Symmetry Breaking in Four-Dimensional 
Lattice Gauge Theories'', {\it Comm. Math. Phys.} {\bf 96}, 251-284 (1984).
\item{[BKL]} J.Bricmont, K.Kuroda and J.L.Lebowitz, ``First Order Phase
Transitions in Lattice and Continuous Systems: Extension of Pirogov-Sinai
Theory'', {\it Comm. Math. Phys.} {\bf 101}, 501-538 (1985).
\item{[BS]} J.Bricmont and J.Slawny, ``Phase Transitions in Systems with a
Finite Number of Dominant Ground States'', {\it J. Stat. Phys.} {\bf 54},
89-161 (1989).
\item{[DS]} E.I.Dinaburg and Ya.G.Sinai, ``Contour Models with Interaction
and their Applications'', {\it Sel. Math. Sov.} {\bf 7}, 291-315 (1988).
\item{[D]} R.L.Dobrushin, ``Gibbs State Describing Phase Coexistence for
Three Dimensional Ising Model'', {\it Theor. Probab. and Appl.} N4, 619-639
(1972). 
\item{[DP]} R.L.Dobrushin and E.A.Pecherski, ``Uniqueness Conditions for
Finitely Dependent Random Fields'', in {\it Colloquia mathematica
Societatis Janos Bolyai}, {\bf 27}, {\it Ran\-dom Fi\-elds}, {\bf 1},
J.Fritz, J.L.Lebowitz and D.Szasz, eds., Amsterdam, New York:
North-Holland Pub., 223-262 (1981).   
\item{[DZ]} R.L.Dobrushin and M.Zahradnik, ``Phase Diagrams for 
Continuous Spin Models. Extension of Pirogov-Sinai Theory'', in {\it
Mathematical Problems of Statistical Mechanics and Dynamics}, R.L.Dobrushin
ed., Dordrecht, Boston: Kluwer Academic Publishers, 1-123 (1986).
\item{[HKZ]} P.Holicky, R.Kotecky and M.Zahradnik, Rigid Interfaces for
Lattice Models at Low Temperatures'', {\it J. Stat. Phys.} {\bf 50},
755-812 (1988).
\item{[LM]} J.L.Lebowitz and A.E.Mazel, ``A Remark on the Low Temperature
Behavior of the SOS Interface in Halfspace'', {\it
J. Stat. Phys.}, {\bf 84}, 379-391 (1996).
\item{[LML]} J.L.Lebowitz and A.Martin-L\"of, ``On the uniqueness of the
equilibrium state for Ising spin systems'', {\it Comm. Math. Phys.} {\bf
25}, 276-282 (1971).
\item{[M1]} D.G.Martirosyan, ``Uniqueness of Gibbs States in Lattice Models
with One Ground State'', {\it Theor. and Math. Phys.} {\bf 63}, N1,
511-518 (1985).
\item{[M2]} D.G.Martirosyan, ``Theorems Concerning the Boundary Layers in
the Classical Ising Models'', {\it Soviet J. Contemp. Math. Anal.} {\bf 22},
N3, 59-83 (1987).
\item{[P]} Y.M.Park, ``Extension of Pirogov-Sinai
Theory of Phase Transitions to Infinite Range Interactions I. Cluster
Expansion and II. Phase Diagram'', {\it Comm. Math. Phys.} {\bf 114},
187-218 and 219-241 (1988).
\item{[PS]} S.A.Pirogov and Ya.G.Sinai, ``Phase Diagrams of Classical
Lattice Systems'', {\it Theor. and Math. Phys.} {\bf 25}, 358-369,
1185-1192 (1975).
\item{[Se]} E.Seiler, ``Gauge Theories as a Problem of Constructive
Quantum Field Theory and Statistical Mechanics'', {\it Lect. Notes in
Physics}, {\bf 159}, Berlin: Springer-Verlag (1982).
\item{[Si]} Ya.G.Sinai, {\it Theory of Phase Transitions}, Budapest:
Academia Kiado and London: Pergamon Press (1982).
\item{[Sl]} J.Slawny, ``Low-Temperature Properties of Classical Lattice
Systems: Phase Transitions and Phase Diagrams'', in {\it Phase Transitions
and Critical Phenomena} {\bf 11}, C.Domb and J.L.Le\-bo\-witz, eds.,
Oxford: Pergamon Press, 128-205 (1987).
\item{[Sh]} S.B.Shlosman, ``Uniqueness and Half-Space Nonuniqueness of
Gibbs States in Czech Models'', {\it Theor. and Math. Phys.} {\bf 66},
284-293, 430-444 (1986).
\item{[Z]} M.Zahradnik, ``An Alternate Version of Pirogov-Sinai
Theory'', {\it Comm. Math. Phys.} {\bf 93}, 559-581 (1984).

\end